**Baseline-free Damage Detection and Localization on Composite Structures with Unsupervised Kolmogorov-Arnold Autoencoder and Guided Waves**

Yunlai Liao[1, 2], Yihan Wang[1], Chen Fang[2], Xin Yang[2], Xianping Zeng[3],
Dimitrios Chronopoulos[2*], Xinlin Qing[1*]

1. School of Aerospace Engineering, Xiamen University, Xiamen, Fujian 361005, China
2. Department of Mechanical Engineering & Division of Mechatronic System Dynamics (LMSD), KU Leuven, Ghent, 9000, Belgium
3. Fujian Key Laboratory of Intelligent Processing Technology and Equipment, Fujian University of Technology, Fuzhou, Fujian 350118, China

---

**Abstract**

Structural health monitoring (SHM) ensures the safety and longevity of structures such as aerospace equipment and wind power installations. Developing a simple, highly flexible, and scalable SHM method that does not depend on baseline models is significant for ensuring the operational integrity of advanced composite structures. In this regard, a hybrid baseline-free damage detection and localization framework incorporating an unsupervised Kolmogorov-Arnold autoencoder (KAE) and modified probabilistic elliptical imaging algorithm (MRAPID) is proposed for damage detection and localization in composite structures. Specifically, KAE was used to process the guided wave signals (GW) without any prior feature extraction process. The KAE continuously learns and adapts to the baseline model of each structure, learning from the response characteristics of its undamaged state. Then, the predictions from KAE are processed, combined with the MRAPID to generate a damage probability map. The performance of the proposed method for damage detection and localization was verified using the simulated damage data obtained on wind turbine blades and the actual damage data obtained on composite flat plates. The results show that the proposed method can effectively detect and localize damage and can achieve multiple damage localization. In addition, the method outperforms classical damage detection algorithms and state-of-the-art baseline-free damage detection and localization methods in terms of damage localization accuracy.

**Key words**: Baseline-free, Autoencoder, Kolmogorov-Arnold networks, Guided wave, Damage detection and localization

---

[*]Corresponding author: Xinlin Qing, E-mail address: xinlinqing@xmu.edu.cn
Dimitrios Chronopoulos, E-mail address: dimitrios.chronopoulos@kuleuven.be

## 1. Introduction

Composite materials are widely used in aerospace structures and wind turbine blades due to their high specific strength, stiffness, and lightweight properties [1]. However, with increased usage time, internal damages such as delamination or debonding may occur in composite structures, posing serious safety risks to the equipment [2]. Therefore, structural health monitoring and damage detection for composite structures are of critical importance [3].

To ensure the safety and reliability of composite structures, various non-destructive evaluation techniques, such as thermography [2, 4], X-ray [5], eddy current [6], and vibrational testing [7], have been developed and applied for damage detection in composites. However, the effectiveness of these techniques heavily depends on the operator's skills and experience [8]. In contrast, guided wave (GW)-based Structural Health Monitoring (SHM) techniques offer a more promising alternative with elliptical probability imaging (PAPID) and delay-and-sum imaging (DAS) algorithms standing out in particular [9, 10]. However, these algorithms rely on calculating robust damage index (DI) and accurate times of flight (TOF) from both baseline signals and current signals [11, 12]. However, baseline data is not always available, which can easily lead to incorrect diagnoses of structural health conditions [13].

Consequently, there is growing interest in developing effective baseline-free techniques that do not require baseline models of the monitored structure [14]. Instead, they utilize advanced signal processing techniques to assess the condition of the monitored structure [15]. Typically, baseline-free techniques can be categorized into physical model-based methods, signal feature extraction methods, and machine learning-driven methods [16]. Among them, physical model-based methods rely on the physical characteristics of GW propagation and utilize wavefield distortions caused by damage for localization. Typical methods include time reversal [17], virtual time reversal [18], nonlinear frequency modulation [19]. Signal feature extraction methods employ real-time or online-constructed dynamic reference baselines to detect and localize damage by comparing differences between current signals and dynamic baselines, such as instantaneous baseline [20, 21]. Machine learning-driven methods learn intrinsic representations of the undamaged structural state through machine learning approaches (e.g., autoencoder [22], dictionary learning [23], K-means [24]), using reconstruction errors as damage indexes (DI) to achieve localization. Physical model-based methods and signal feature extraction methods require assumptions about GW propagation, material properties of the structure, or sensor arrangements [15, 25]. However, in practical damage detection tasks, these physics-based assumptions are only approximately valid. This limitation may lead to false alarms during damage detection and localization processes. Therefore, researchers have gradually begun to shift the focus on research towards baseline-free damage detection methods driven by machine learning approaches.

Currently, a limited amount of research has been conducted on machine learning-based baseline-free damage detection. Recent work by Mazzi et al. [22] proposed a baseline-free damage localization framework combining convolutional autoencoder with RAPID, which effectively achieved damage localization in composite structures. However, this method requires training a unique convolutional autoencoder for each sensing path, necessitating extensive baseline signal collection for each model, making it undoubtedly time-consuming and labor-intensive. Another limitation is its dependence on expert knowledge to obtain the wave velocity of the monitored structure, where velocity inaccuracies can reduce localization precision. Zhu et al. [23] integrated dictionary learning with the RAPID algorithm to propose a baseline-free damage detection and localization method. Although this method achieved good damage localization accuracy, it still has certain limitations. Firstly, acquiring the first direct wave of GW may not always be feasible in real-world composite structures due to their complex geometries [26]. Secondly, even after obtaining virtual baselines, expert knowledge is still required to calculate DI values for each sensing path, and obtaining robust DIs remains a significant challenge due to DI representation diversity. Barzegar et al. [24] developed a baseline-free damage imaging method for composite lap joints using K-means clustering to identify anomalies and assign damage coefficients. However, this method assumes completely consistent material properties along sensing paths of the same orientation. Furthermore, existing studies mainly focus on single-damage scenarios and have not extended to multiple-damage detection tasks. Most critically, no studies have validated the applicability of current baseline-free techniques for large-scale complex composite structures like wind turbine blades, significantly limiting their generalizability.

Given the above limitations, a baseline-free damage detection and localization method based on unsupervised Kolmogorov-Arnold autoencoder (KAE) and modified RAPID (KAE-MRAPID) is proposed, aiming to develop an efficient end-to-end damage detection and localization technique for large and complex composite materials. Compared to existing baseline-free methods based on unsupervised autoencoder [22], this method supports transferability, meaning it maintains excellent damage detection and localization performance even when applied to different structures, without the need to train unique models for each sensing path. In comparison to literature [23], the proposed method is fully automated and the DIs are computed directly by the KAE without the need to rely on expert knowledge to manually compute the DIs as in the literature [23]. When monitoring large-scale structures, such as wind turbine blades, the integrated sensor network self-assembly method can reconstruct a small monitoring network based on the damage location, enabling multi-damage detection and localization. To validate the effectiveness of the proposed KAE-MRAPID model, comparative experiments were conducted on wind turbine blades and composite plates using traditional RAPID and existing baseline-free damage detection models. The results show that the proposed KAE-MRAPID achieves the highest localization accuracy while providing relatively competitive operational efficiency. In summary, this study makes significant contributions in the following aspects:

1) A fully unsupervised end-to-end SHM method for composites is proposed that allows the detection and localization of single or multiple damages in large area composite structures;

2) This study is the first attempt to integrate the Kolmogorov-Arnold network as autoencoder into a baseline-free damage detection and localization method for SHM in large-area wind turbine composites;

3) The proposed method does not rely on any material properties or signal processing knowledge, demonstrating its potential for application to different structures with retraining only on undamaged signals;

4) The proposed KAE-MRAPID framework was validated using simulated damage data and actual damage data to demonstrate the potential and effectiveness of applying the KAE in the field of SHM for baseline-free damage detection and localization.

The remainder of this paper is organized as follows. Section 2 introduces the proposed baseline-free damage diagnosis and localization framework. Section 3 describes the experimental details and specific results of two case studies. Finally, Section 4 provides a summary of this paper.

## 2. Baseline-free damage detection and localization framework

This paper proposes a baseline-free method based on KAE and MRAPID to detect and localize damage in composite structures, as shown in Fig. 1. The main principle of the method is to construct virtual baseline signals for the detected structure using KAE. One of the key advantages of the proposed approach is that it eliminates the need for training on GW signals from damaged structures, as collecting such GW signals from real-world structures is often unfeasible. Another significant advantage is that the DI values in the damage imaging process are automatically identified and calculated by the KAE, without the need for manual searching, overcoming the previous limitation of having to use trial and error to select the optimal DI. The proposed method uses a PZT array installed on the structure to excite and sense GW. The raw GW data is normalized to the range [0, 1], and then the KAE is trained purely using the signals from the undamaged state of the structure. For large-scale monitoring areas, the region is divided into several smaller areas. The trained KAE is then used to reconstruct the error distribution for the health state of each small area, and the Health Index (HI) of the small area structure is obtained using the quantile method of probabilistic statistics. During damage monitoring, the KAE is used to process data from PZT to reconstruct the virtual baseline signal. The loss function is used to obtain the DI values for all sensor paths in each region, and the DIs for each region are further obtained through quantiles. By comparing the health indices and DIs of each region, the damaged area is determined. Subsequently, the DI values obtained from the KAE and the MRAPID algorithm are used to achieve precise damage localization.

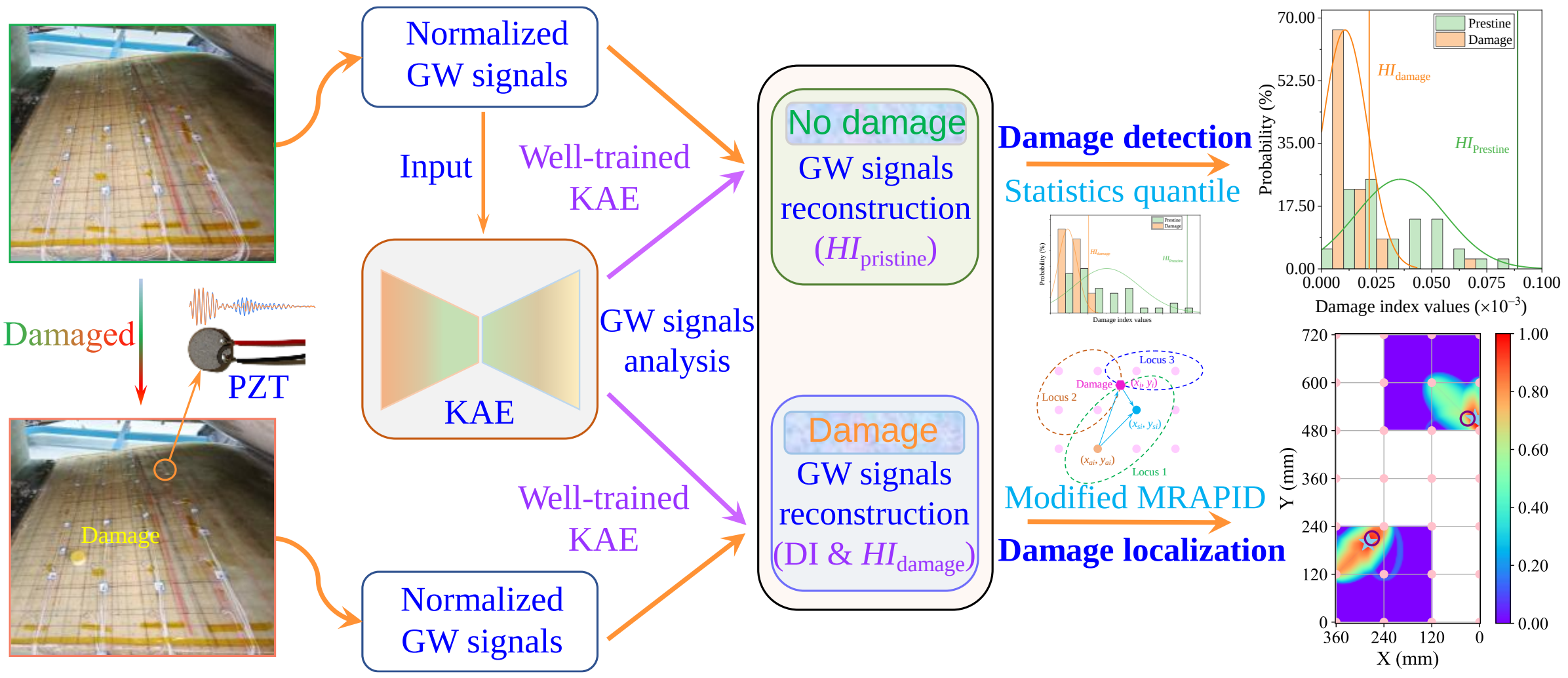

Fig. 1. **Overview of the hybrid baseline-free damage detection and localization framework.**

### 2.1. The proposed Kolmogorov-Arnold autoencoder

The KAN network takes the Kolmogorov-Arnold theorem as its theoretical foundation. The ability to fit the model and flexibly represent multivariate continuous functions is improved by using learnable activation functions and using spline functions instead of traditional weight parameters. For a given input vector $x \in R^n$, the general representation [27] of the KAN is

$$\mathrm{KAN}(x) = \left(\mathbf{\Phi}_{L-1} \circ \mathbf{\Phi}_{L-2} \circ \cdots \circ \mathbf{\Phi}_1 \circ \mathbf{\Phi}_0\right) x \,, \tag{1}$$

where $\Phi_{L-1}$ is the function matrix corresponding to the $L$-1st KAN layer.

In this paper, a novel KAE autoencoder based on KAN networks is proposed for reconstructing virtual baseline signals. As far as we know, this is the first time that a KAN network has been used to build an autoencoder for virtual baseline signal reconstruction in SHM. The architecture of the KAE network consists of an encoder and a decoder, as shown at the top of Fig. 2.

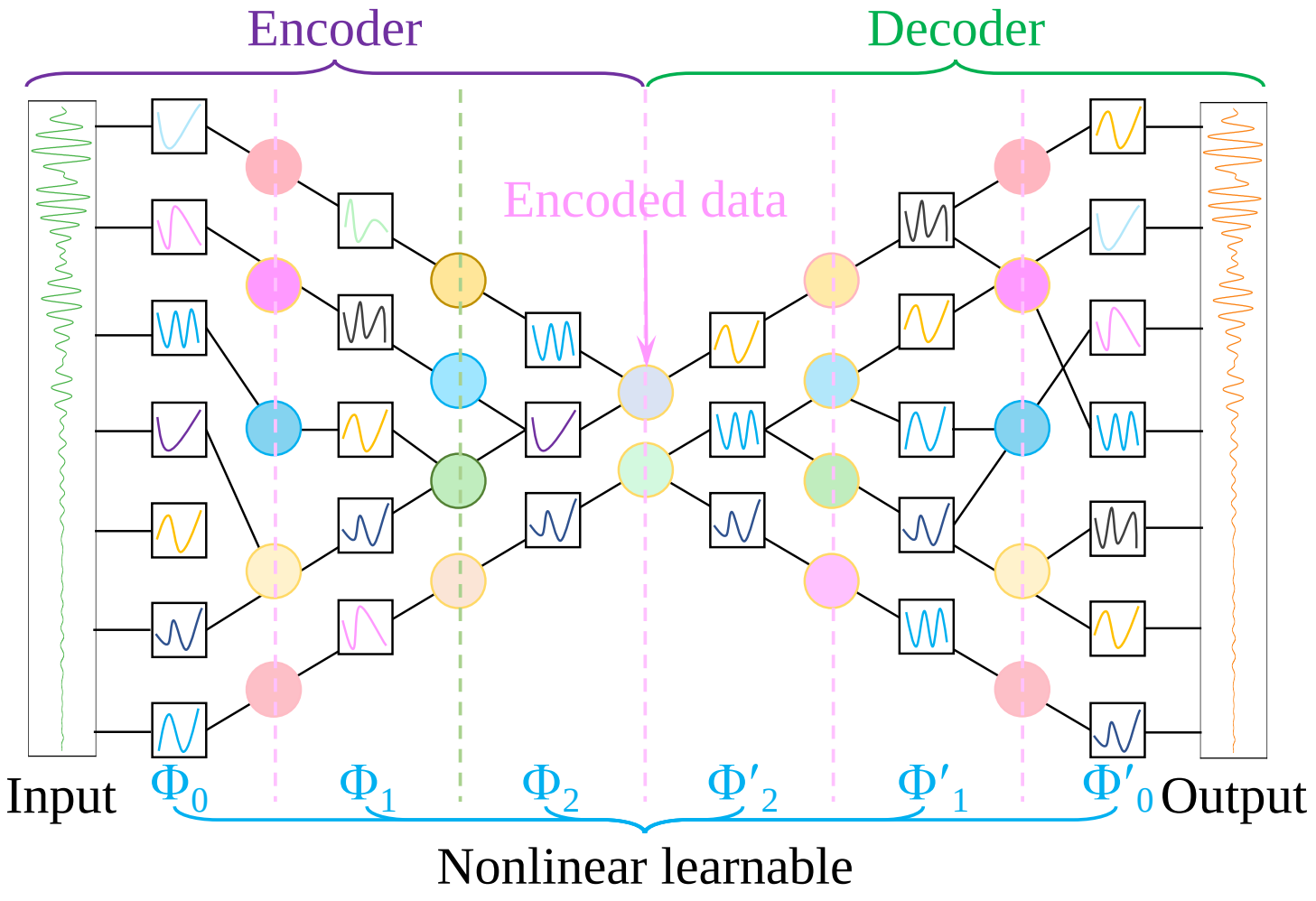

Fig. 2. **Layout of a representative Kolmogorov-Arnold autoencoder.**

### 2.1.1. Encoder

In the encoder, the number of neurons in each layer is reduced sequentially. That is, the input vector ($x_i \in R^m$) is compressed into $l$ ($l < m$) neurons, which constitute the bottleneck layer. The process can be expressed as:

$$S_i = f_{\mathbf{\Phi}}(x) = \left(\mathbf{\Phi}_0 \circ \mathbf{\Phi}_1 \circ \mathbf{\Phi}_2\right) x \,, \tag{2}$$

where the parameter set of the encoder $\theta$ is [$\Phi_0$, $\Phi_1$, $\Phi_2$]. The input vectors are encoded as low dimensional vectors due to the decrease in the number of nodes per layer.

### 2.1.2. Decoder

In the decoder, the number of neurons per layer is increased. This means that the resulting hidden representation $\boldsymbol{s}$ is decoded back into the original space $R^m$. This process can be represented as:

$$x' = f_{\mathbf{\Phi}'}(x) = \left(\mathbf{\Phi}_2' \circ \mathbf{\Phi}_1' \circ \mathbf{\Phi}_0'\right) s \,, \tag{3}$$

where the parameter set of the encoder $\theta'$ is [$\Phi_2'$, $\Phi_1'$, $\Phi_0'$].

### 2.1.3. Loss function

The Adam optimization algorithm [28] was used to optimize the parameters within the model in order to minimize the reconstruction error between the inputs and outputs of the model:

$$\underset{\theta,\theta'}{\operatorname{argmin}} \frac{1}{N}\sum_{i=1}^{N}\mathcal{L}\left(\mathbf{x}_i, \mathbf{x}_i'\right) = \underset{\theta,\theta'}{\operatorname{argmin}} \frac{1}{N}\sum_{i=1}^{N}\mathcal{L}\left(\mathbf{x}_i, f_{\theta'}\left(f_\theta\left(\mathbf{x}_i\right)\right)\right), \tag{4}$$

where $\mathcal{L}\left(\mathbf{x}_i, \mathbf{x}_i'\right) = \sum_{j=1}^{m}(x_{ij} - x_{ij}')^2$ represents the reconstruction error (Mean Squared Error, MSE).

The goal is to make the output signal as close as possible to the original input signal. Details of the encoder and decoder networks are given in Table 1. In the encoder network, the number of neurons in each layer from the first to the last are 6000, 512, 256, and 8, respectively. The number of neurons in each layer of the decoder network is inversely proportional to the corresponding layer in the encoder network.

Table 1. Parameters for training autoencoder network

| Parameters | Values in each layer |
|---|---|
| Neuron number of layers in encoder network | 6000, 512, 256, 8 |
| Neuron number of layers in decoder network | 8, 256, 512, 6000 |

### 2.2. Damage detection and localization

Assuming that a structure is monitored by $n$ transducers in a pitch-catch manner, the total number of transducer paths $N$=$n$×($n$-1)/2 when the inverse paths are ignored (i.e., $A_1$-$S_2$ and $A_2$-$S_1$ are taken to be $A_1$-$S_2$ only. $A_1$-$S_2$ indicates that the $S_1$ sensor actuate the signal, and the $S_2$ sensor receives the signal). At the same time, the DI of the transducer paths $i$ ($i$ = 1,2,··$N$) is denoted by

$$\mathrm{DI_i} = \mathcal{L}(\mathrm{signal}_i, \mathrm{KANAutoencoder}(\mathrm{signal}_i)) \,, \tag{5}$$

where $\mathcal{L}(\cdot, \cdot)$ is the reconstruction error of the KAE, and $\mathrm{signal}_i$ is the GW signal of the $i$-th sensing path. It can be seen that the DI value is completely calculated automatically by KAE, eliminating the previous limitation of relying on the expert's empirical knowledge to calculate the DI value.

To reflect the health of the monitored structure, a *HI* [16] is created by fusing the DIs, which can be expressed as

$$HI = \mathrm{DI}_{95\%} \,, \tag{6}$$

where $\mathrm{DI}_{95\%}$ indicates the 95% quantile of the *DI* value.

In the case of damage detection, the detection thresholds for *HI* are crucial for a good differentiation between the healthy and damaged state of the structure. When using baseless techniques, the thresholds need to be calibrated beforehand by means of an experimental baseline, since these data are not available during the monitoring process [29]. The calibration is done by feeding the raw signals into the baseless method and thus calculating the corresponding *HI* values. In this article, the damage threshold is set equal to the *HI*, i.e.

$$ThrV = HI \,. \tag{7}$$

According to the definition of Eq.(7), damage is considered to occur inside the monitored structure if the $HI_{damage}$>$ThrV$ of the damaged structure and vice versa.

Once it is determined that damage has occurred, damage localization is achieved using the RAPID method. RAPID calculates DIs for each actuator-sensor path to identify the probability of damage at the coordinates in the structure. Fig. 3 illustrates the imaging of a single transducer path, where the points A, S, P individually represents the actuator, the sensor and the imaging pixel.

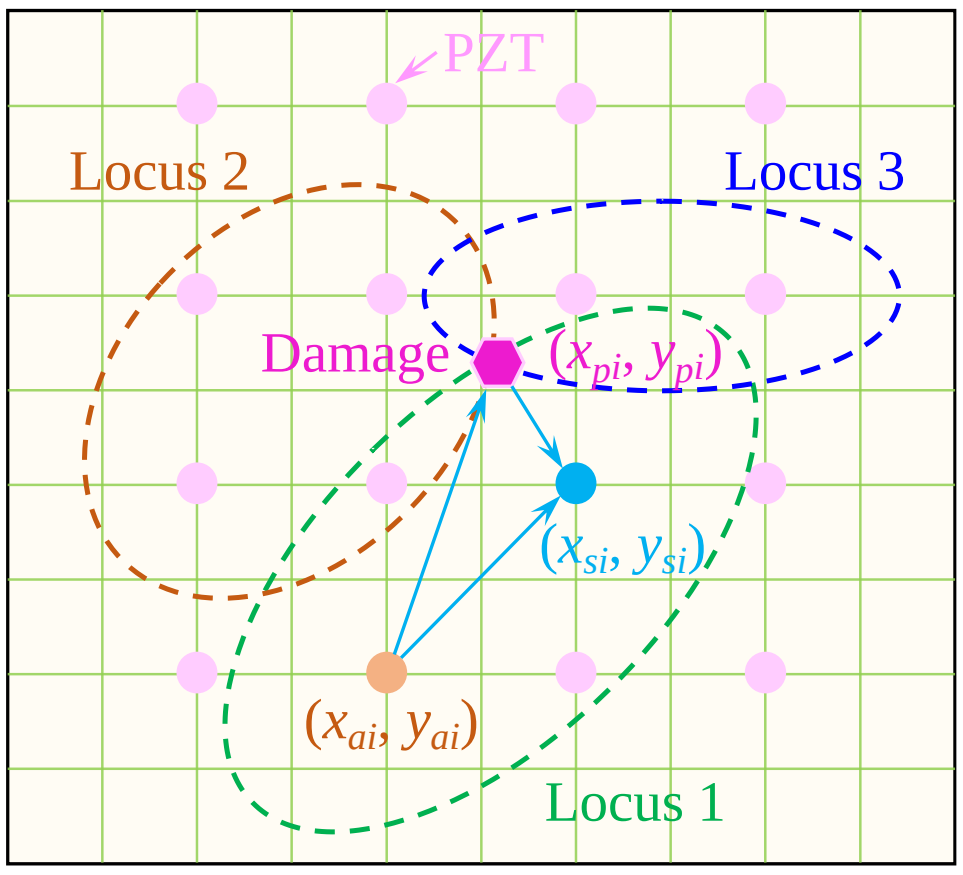

Fig. 3. **Description of damage localization based on RAPID.**

Although RAPID is easy to implement, it also provides satisfactory performance and over the years researchers have proposed a variety of improvements to make it better. Zhu et al. [30]. proposed a damage imaging algorithm with a modified probability weighting algorithm that varies linearly with the distance between the actuator and the location where the damage is likely to occur. According to the basic principle of the algorithm, the weight distribution function [30] of the $i$-th excitation-sensing channel, can be expressed as

$$W_i\left[d_i(x_i,y_i)\right]=\begin{cases}1+\left[\dfrac{d_i(x_i,y_i)}{r}-\left(1-\dfrac{DI_i}{DI_{max}}\right)\right], & d_i(x_i,y_i)<\left(1-\dfrac{DI_i}{DI_{max}}\right)\cdot r\\ 1-\left[\dfrac{d_i(x_i,y_i)}{r}-\left(1-\dfrac{DI_i}{DI_{max}}\right)\right], & \left(1-\dfrac{DI_i}{DI_{max}}\right)\cdot r\leqslant d_i(x_i,y_i)<\left[1+\left(1-\dfrac{DI_i}{DI_{max}}\right)\right]\cdot r\\ 0, & \text{other}\end{cases}, \tag{8}$$

where $W_i[d_i(x_i, y_i)]$ is the probability distribution function of the $i$-th sensor, $DI_{max}$ represents the maximum DI among all excitation-sensing channels, $r$ denotes the sensing region radius of the PZT surface-mounted on the structure, and $d_i(x, y)$ is the Euclidean distance between the pixel point $(x, y)$ and the path of the $i$-th excitation-sensing channel, which can be be expressed as

$$d_{ij}(x,y)=\frac{\sqrt{(x-x_i)^2+(y-y_i)^2}+\sqrt{(x-x_j)^2+(y-y_j)^2}}{\sqrt{(x_i-x_j)^2+(y_i-y_j)^2}}, \tag{9}$$

where ($x_i$, $y_i$) and ($x_j$, $y_j$) are the coordinates of actuator $i$ and sensor $j$, respectively.

Finally, damage imaging in the monitored structure can be achieved by plotting the fusion response of each pixel point ($x_i$, $y_i$) in the structure, i.e.

$$P_i(x_i, y_i) = \sum_{i=1}^{N}\left(W_i\left[d_i(x_i, y_i)\right]\cdot \mathrm{DI}_i\right), \tag{10}$$

where $N$ is the total number of PZTs in the sensing network.

After obtaining the predicted damage, the performance of the proposed method needs to be evaluated using appropriate metrics. Table 2 outlines the commonly utilized tools for model evaluation, where lower values of Root Mean Square Error (RMSE), Mean Absolute Percentage Error (MAPE), and Mean Absolute Error (MAE) signify better model performance [31, 32].

Table 2. Details of the evaluation metrics

| Metric | Equation |
|---|---|
| RMSE | $\mathrm{RMSE} = \sqrt{\frac{1}{n}\sum_{i=1}^{n}(\hat{y}_i - y_i)^2}$ |
| MAPE | $\mathrm{MAPE} = \frac{100\%}{n}\sum_{i=1}^{n}\left\lvert\frac{\hat{y}_i - y_i}{y_i}\right\rvert$ |
| MRE | $\mathrm{MRE} = \frac{1}{n}\sum_{n}^{i=1}[\sqrt{\left(x_{\mathrm{T}}^{i} - x_{\mathrm{P}}^{i}\right)^2 + \left(y_{\mathrm{T}}^{i} - y_{\mathrm{P}}^{i}\right)^2}/L]\times 100\%$ |

### 2.3. A multi-damage localization method via self-assembly of sensor networks

The proposed multi-damage localization method based on self-assembly of sensing networks is inspired by the DI consolidation algorithm proposed by Qiu et al [33]. The core of the algorithm is to divide a large composite area into multiple sub-regions, and then calculate the average combined DI for each sub-region. Subsequently, the damage detection and localization algorithm are executed for the sub-regions whose average composite DI exceeds a preset threshold, until the detection and localization of the damage of the whole structure is achieved. The advantage of this approach is that it significantly improves the efficiency of large-scale composite damage imaging.

The proposed algorithm does not require the setting of damage thresholds compared to the method proposed by Qiu et al [33]. It is directly obtained by the KAE autoencoder and the quantile method of probability theory, which overcomes the limitation of manually setting the damage thresholds for each region manually. The pseudo-code of the proposed multi-damage localization method based on self-assembly of sensor networks is shown in Algorithm 1.

**Algorithm 1**

**Input**: Guided wave signals
**Outout**: Damage imaging and damage locations
**Step 1**: Dividing large composite areas into multiple sub-areas
**Step 2**: Obtaining all damages via Eq.(5)-Eq.(10)
**Step 3**: Creating processed_damages list and final_damages list
**Step 4**: for damage in all damages:
**Step 5**: if damage in processed_damages:
Continue
**Step 6**: Determining sub-area where the damage is located using Eq.(5)-Eq.(7)
**Step 7**: Obtaining the overlap_damages
**Step 8**: For overlap_damage in overlap_damages:
**Step 9**: Calculating the Euclidean distance between the current damage and the untreated damage
**Step 10**: if Euclidean distance < threshold:
**Step 11**: Considering them as identical damage and saving their average in final_damages, with the second point stored in processed_damages
**Step 11**: else:
**Step 12**: Preserving current damage in final_damages
**Step 13**: Repeating all the damages and imaging the damages in final_damages

## 3. Case studies

The performance of the unsupervised damage localization method proposed in this work is validated by two case studies: an experimentally validated GW dataset obtained on a complex wind turbine blade, and another experimental GW dataset obtained on a composite flat plate, where the GW signals on the composite plate is the actual damage data.

### 3.1. Case 1: Application to the complex wind turbine blade structures

One of the structural segments of a wind turbine blade was selected as an experimental subject to verify the effectiveness of the proposed baseline-free damage detection and localization method. The specimen was cut from an intact wind turbine blade, and its main components include the trailing plate, girder cap and front plate, as shown in Fig. 4(a). The specimen consists of a glass fibre reinforced composite (GFRP) and a sandwich structure with a chord length of 2440 mm and a width of 1000 mm. To collect signals on the structural integrity of the wind turbine blade, a piezoelectric sensing network is arranged on the backside of the beam cap structure. The network consists of 28 piezoelectric transducers (PZTs) with a diameter of 8 mm and a thickness of 0.48 mm, which are cured on the backside of the beam cap structure by an AB-type epoxy adhesive, as shown in Fig. 4(b). The thickness of the monitored structure was 38 mm, and its upper and lower structures were GFRP with a foam layer in the middle structure as shown in Fig. 4(c). The detailed locations of the PZTs are shown in Fig. 4(d).

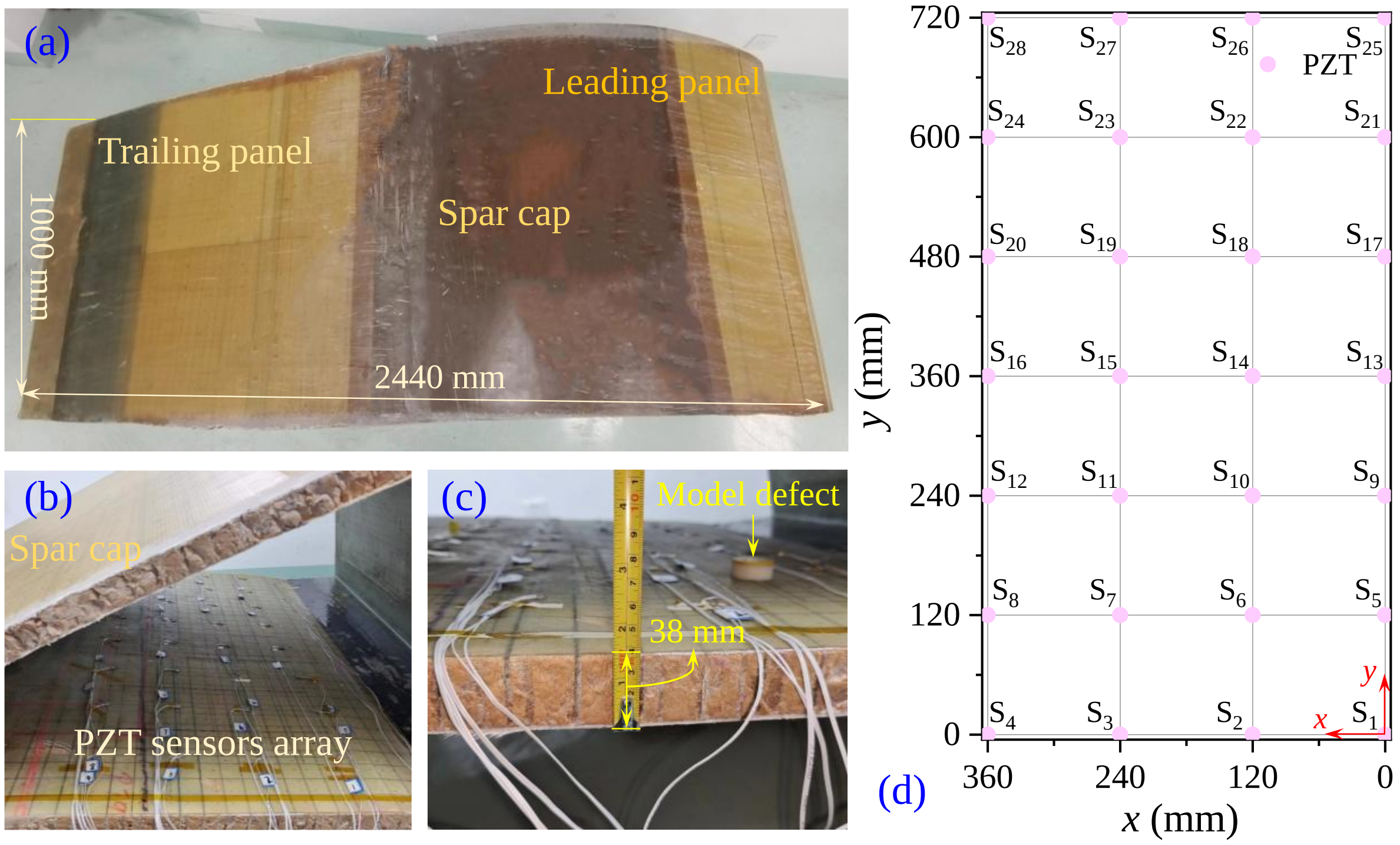

Fig. 4. **Wind turbine blade**: (a) Overall structure of the wind turbine blade, (b) Surface of the structure on which the sensor is mounted, (c) Thickness of the structure to be monitored, (d) Detailed location of the sensor arrangement.

The ultrasonic GW monitoring experimental platform, shown in Fig. 5(a), consists of a computer, an ultrasound detector, and a wind turbine blade specimen. Among them, the computer is used to set the system parameters of the GW signal and display and store the data in real time. The ultrasonic monitoring system is developed by Dalian Junsheng Technology Co., Ltd. and its mainframe integrates signal generation, power amplification, high-speed acquisition, and band-pass filtering. Since the GW signals received by the sensors spaced far apart are very weak, or even cannot be received, the whole monitoring area is divided into six target areas, as shown in Fig. 5(b). The sensor numbers assigned to each target area are shown in Fig. 5(c). During the acquisition of signals, when S1 is excited, only the eight PZTs around it can receive the signals, as shown in the lower right corner of Fig. 5(d). This setup is also applied when S2 to S27 are excited as well, and this setup can speed up the data acquisition. A five-cycle Hanning window signal with an amplitude of 120 V was used as the excitation signal and the sampling rate was set to 12 MHz. Due to the limited resources, it is not practical to collect data on the wind turbine blades by fabricating real damage. Considering that the effect of using wave-absorbing adhesive is similar to the real damage during GW propagation. Therefore, the damage is simulated using the method based on absorbing wave bonding coupling [32], and the simulated damage is shown in the lower left corner of Fig. 5(a). The acquisition process of the GW signals consists of two main steps. First, 43 baseline measurements of the intact structure were recorded. 80% (34 baseline measurements) were used to train the KAE, and the other 20% (9 baseline measurements) were used to determine the *HI* in the absence of damage. Then, the simulated damage was randomly placed on the wind turbine blade and the GW signal was re-excited to capture the GW signal with damage. It should be noted that for single damage, only one simulated damage was placed on the GFRP. For two damages, two damages are placed simultaneously in the monitoring area.

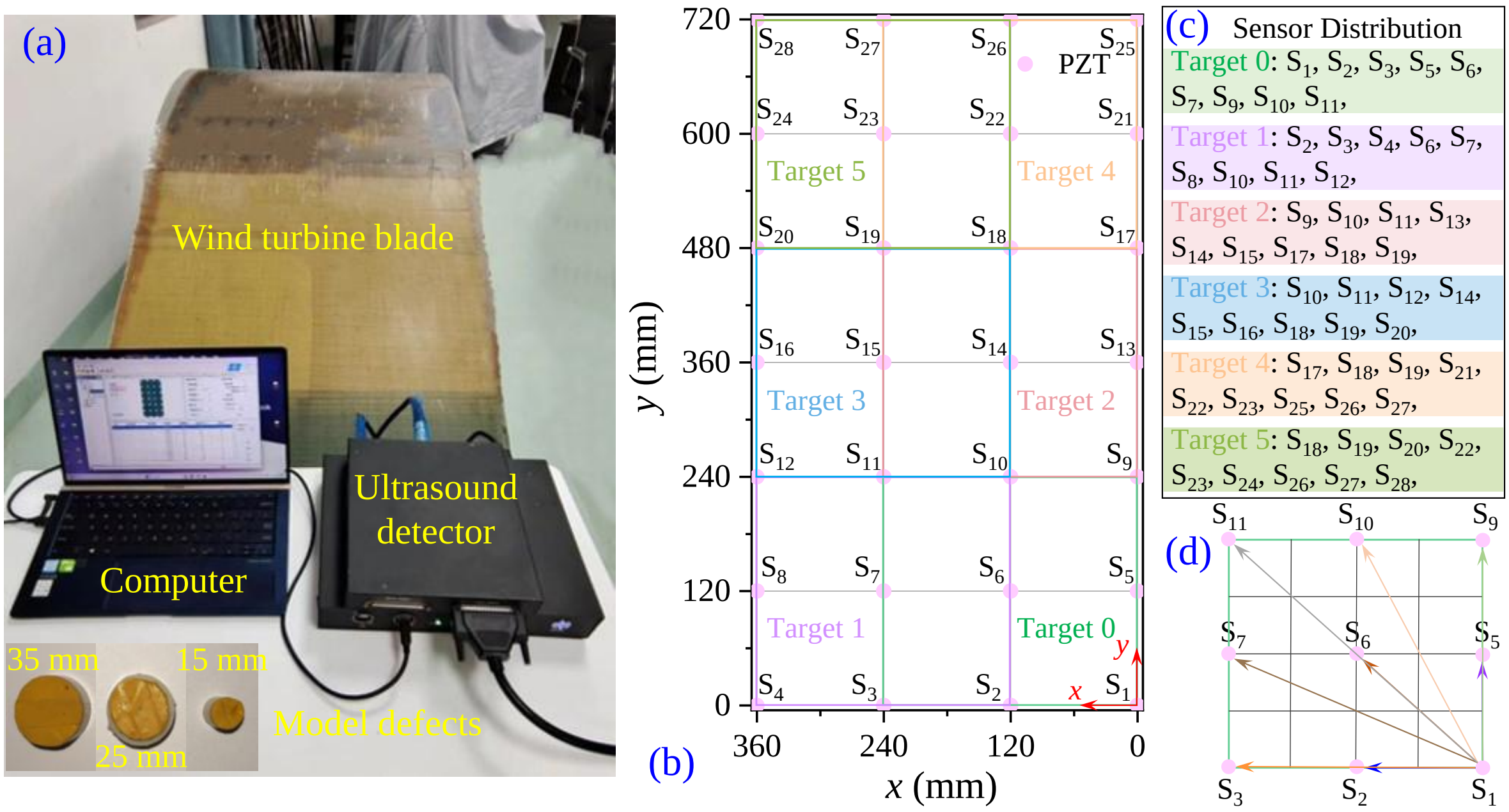

Fig. 5. **Experimental test rigs**: (a) The GW test experimental platform, (b) Detailed map of area division and schematic of signal acquisition, (c) Sensor number contained in each sub-area. (d) Schematic of the sensing path.

### 3.1.1. Training process and damage threshold determination

The baseline-free damage detection and localization method was implemented using Pytorch on a laptop with an R7-5800H CPU and RTX 3060 GPU. The Adam optimizer minimized the MSE between input and reconstructed signals. To reduce randomness, all experiments were repeated three times, with Adam's parameters listed in Table 3. The trained model was evaluated using experimental data, and loss variation during training is shown in Fig. 6. After about 25 iterations, the MSE stabilized, and training and validation losses were similar, indicating no overfitting and effective signal reconstruction.

Table 3. parameters of Adam's optimization algorithm

| Parameter | Learning rate | Batch size | Epoch | Weight decay | gamma |
|---|---|---|---|---|---|
| Valve | 0.001 | 16 | 100 | 0.000001 | 0.95 |

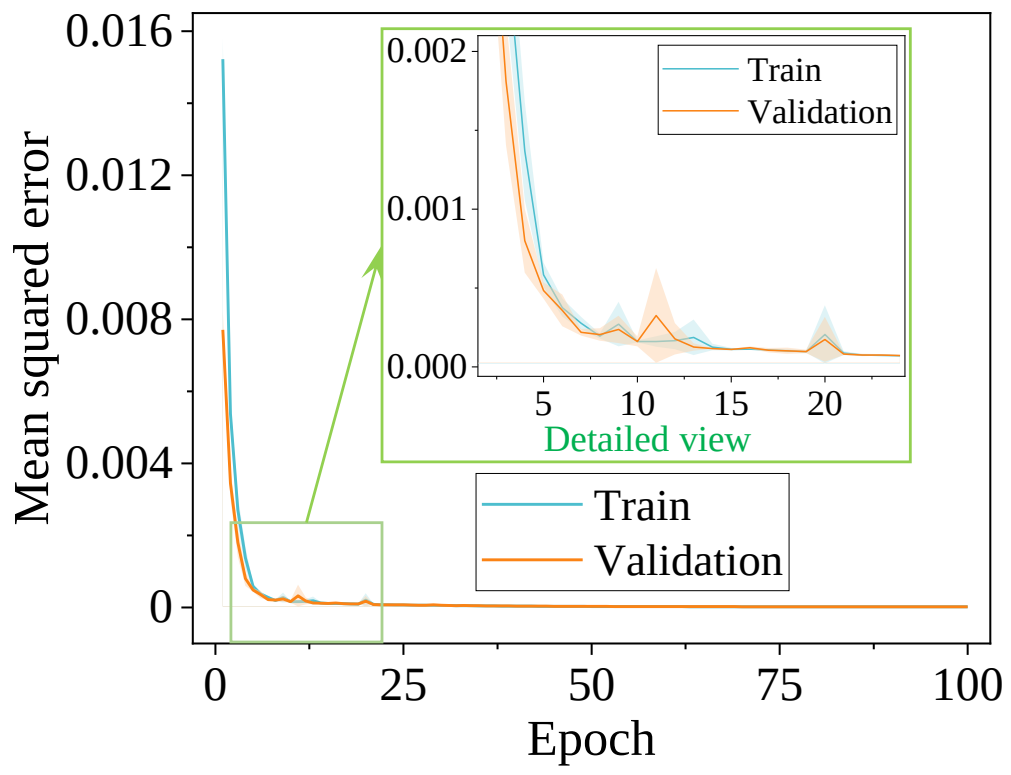

Fig. 6. **Learning curves of the proposed KAE on raw GW signals from wind turbine blades.**

Fig. 7 illustrates the average reconstruction MSE ε for different sensing paths inside different regions under healthy state. The following two conclusions can be derived: 1) The reconstruction errors of the KAE for different sensing paths in different regions are different. This is mainly due to the fact that GFRP is anisotropic and thus the material properties are different in each region, leading to different propagation behavior of the GW signals. 2) The reconstruction errors of the sensing path signals in all regions are less than 10-4, indicating that the proposed KAE can reconstruct the input signals well.

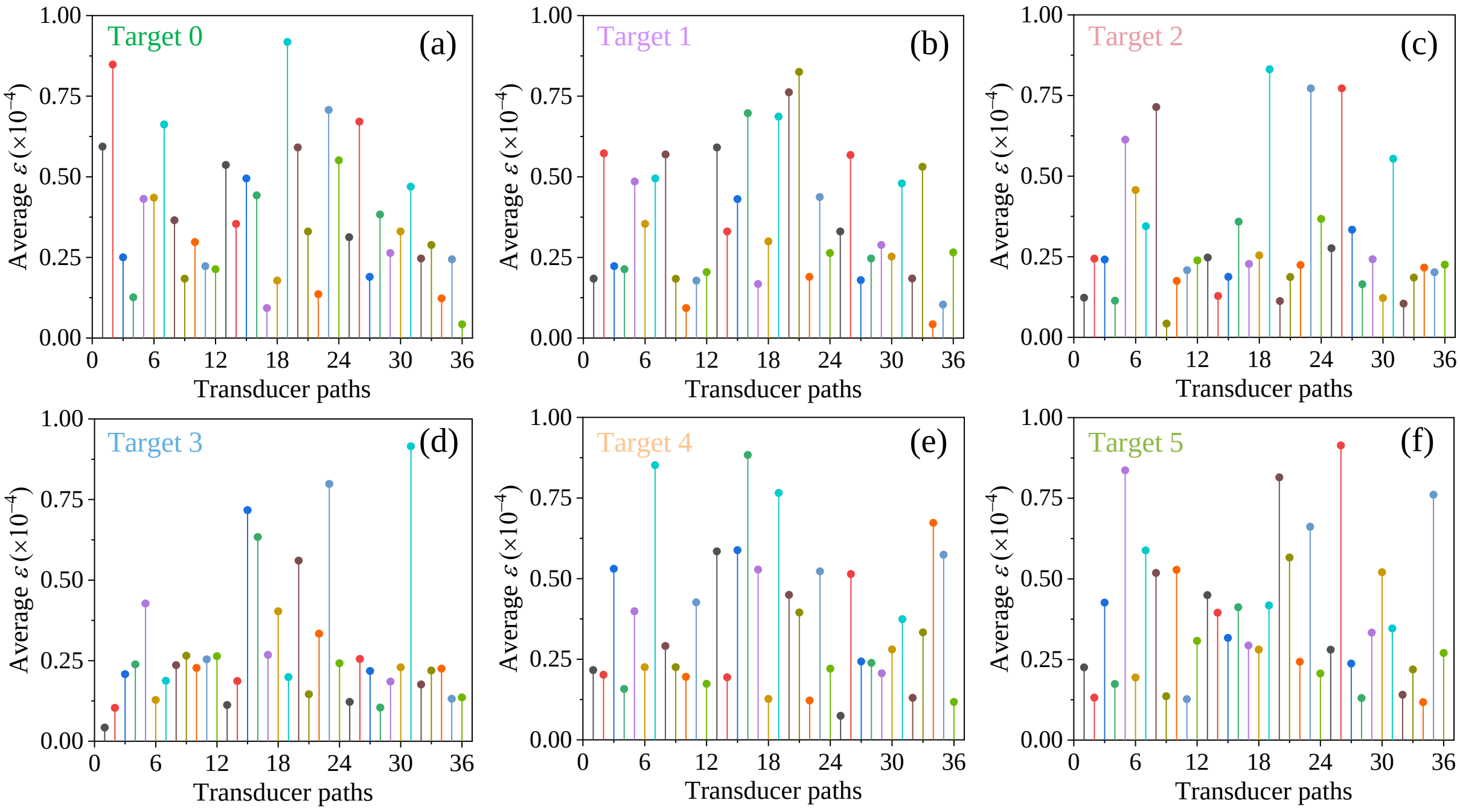

Fig. 7. **Average reconstruction error ε for different target regions**: (a) Target 0, (b) Target 1, (c) Target 2, (d) Target 3, (e) Target 4, (f) Target 5.

After obtaining the reconstructed MSE of all the sensing paths within different regions, the health indices of different monitoring regions within the wind turbine blade under the no-damage condition can be obtained by Eq. (5) and Eq. (6), as shown in Fig. 8. It can be seen that the DI values within different regions show normal distribution in the whole. However, the DI values within different regions are significantly different, which is mainly due to the anisotropy of the GFRP structure.

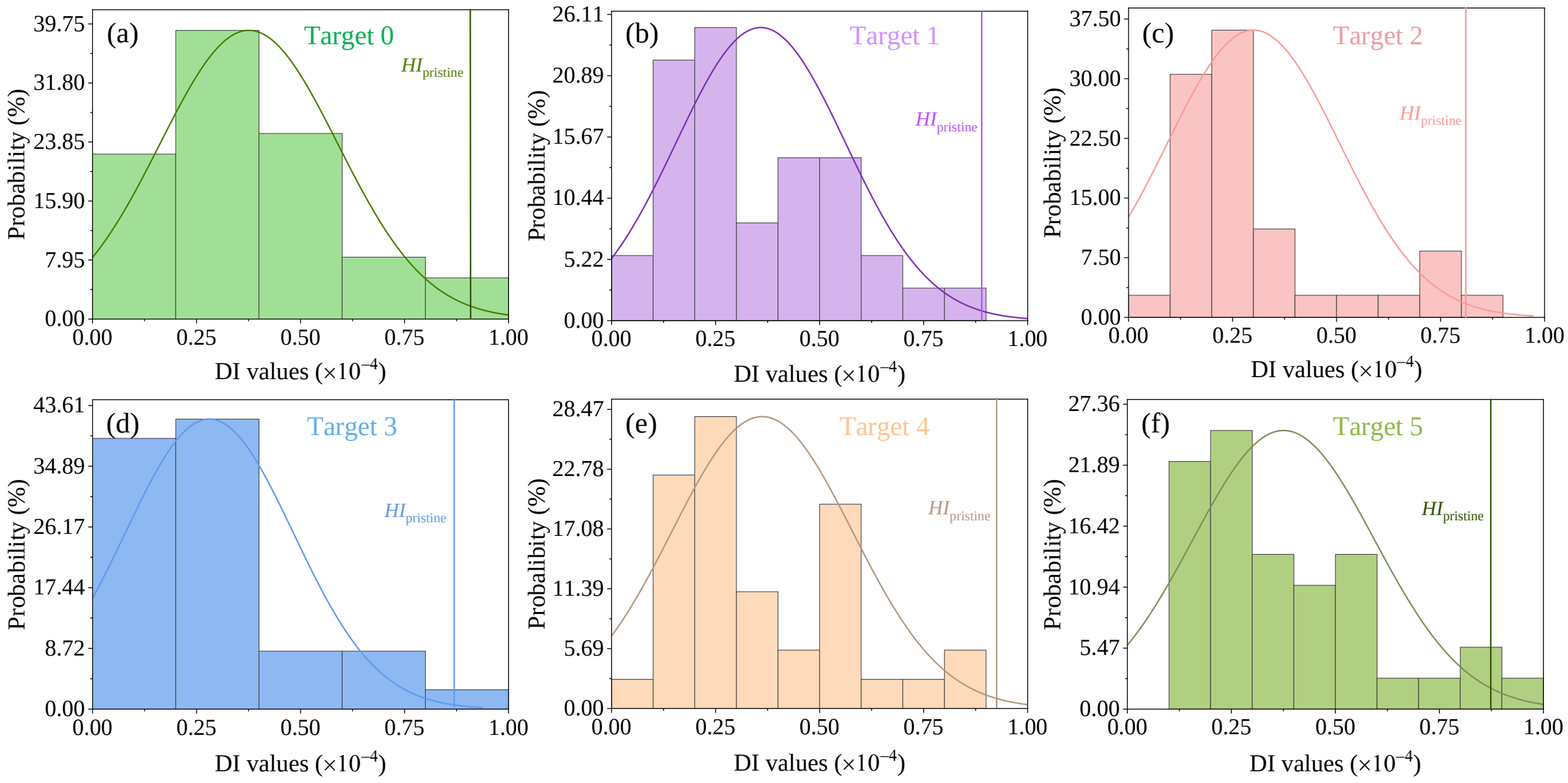

Fig. 8. **Distribution of DI values in the absence of damage in different regions**: (a) Target 0; (b) Target 1; (c) Target 2; (d) Target 3; (e) Target 4; (f) Target 5.

### 3.1.2. Single damage detection and localization

Fig. 9 shows the damage detection and localization results of the proposed method for a single damage scenario. Fig. 9(a)-(f) show the evolution of the *HI* values of the simulated damage under different target regions. In the figures, the *HI*s in the undamaged and damaged states are denoted as $HI_{\text{pristine}}$, and $HI_{\text{damage}}$, respectively. The $HI_{\text{pristine}}$ of each region is calibrated by the pristine signal, which can be calculated by Eq. (5)-(7). It can be seen that the $HI_{\text{damage}}$ value is much higher than the $HI_{\text{pristine}}$ value in the region where damage occurs and vice versa. Fig. 9(g) shows the damage imaging results given by the proposed baseline-free method for the region appearing damage, and it can be seen that the proposed method can effectively localize the damage. Fig. 9(g) shows the DI values of all sensing paths in the region where the damage is located. It is observed that the proposed method can effectively capture the extreme DI values near the damage paths (1-6, 1-2, 1-5) and use these paths to effectively localize the damage.

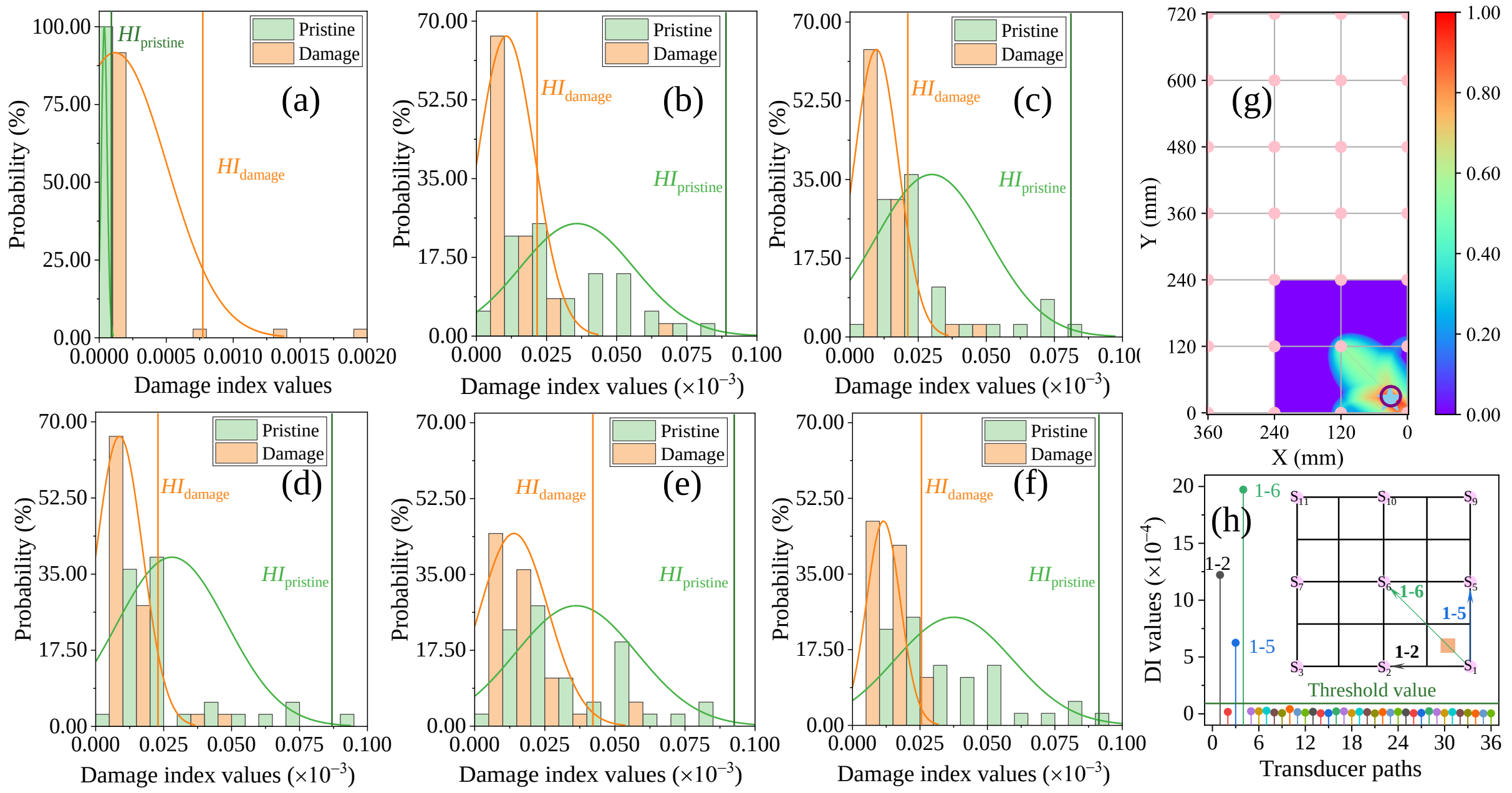

Fig. 9. **(a)-(f) DI distribution in different target regions**: (a) target 0, (b) target 1, (c) target 2, (d) target 3, (e) target 4, (f) target 5; (g) damage imaging map; (h) DI histograms with reciprocal paths and indirect paths ignored. The actual damage is marked with a purple circle, and the predicted damage is represented by a sky blue star (The follow-up is the same and will not be repeated).

In addition, the damage detection and localization performance of the proposed baseline-free method for different damage sizes were further investigated. The distribution of DI values in the region where the damage is located is given in Fig. 10(a, d, g). It can be noted that the $HI_{\text{damage}}$ of the damaged region is always greater than the $HI_{\text{pristine}}$ at different damage sizes. Nevertheless, the curve of the probability density function obtained from the kernel density estimation is gradually shifted to the left as the damage size decreases. This is mainly due to the fact that as the damage decreases, the DIs decrease accordingly, which leads to a right skewed distribution of the probability density function curve. This result is similar to that reported in the literature [23]. Fig. 10(b, e, h) show the damage imaging results for different damage sizes at the same damage location. It is evident that the target region where the damage is located is well evaluated and is not affected by the variation of the damage size. Fig. 10(c, f, i) illustrate the DI values in the damage region, and it can be noticed that most of the paths have DIs smaller than the damage threshold values, and these paths are close to the location of the damage. This finding suggests that the damage information is confined to a few sensor paths, while most of the paths are considered intact, a finding that will motivate the development of self-selecting path algorithms to improve the operational efficiency of the model.

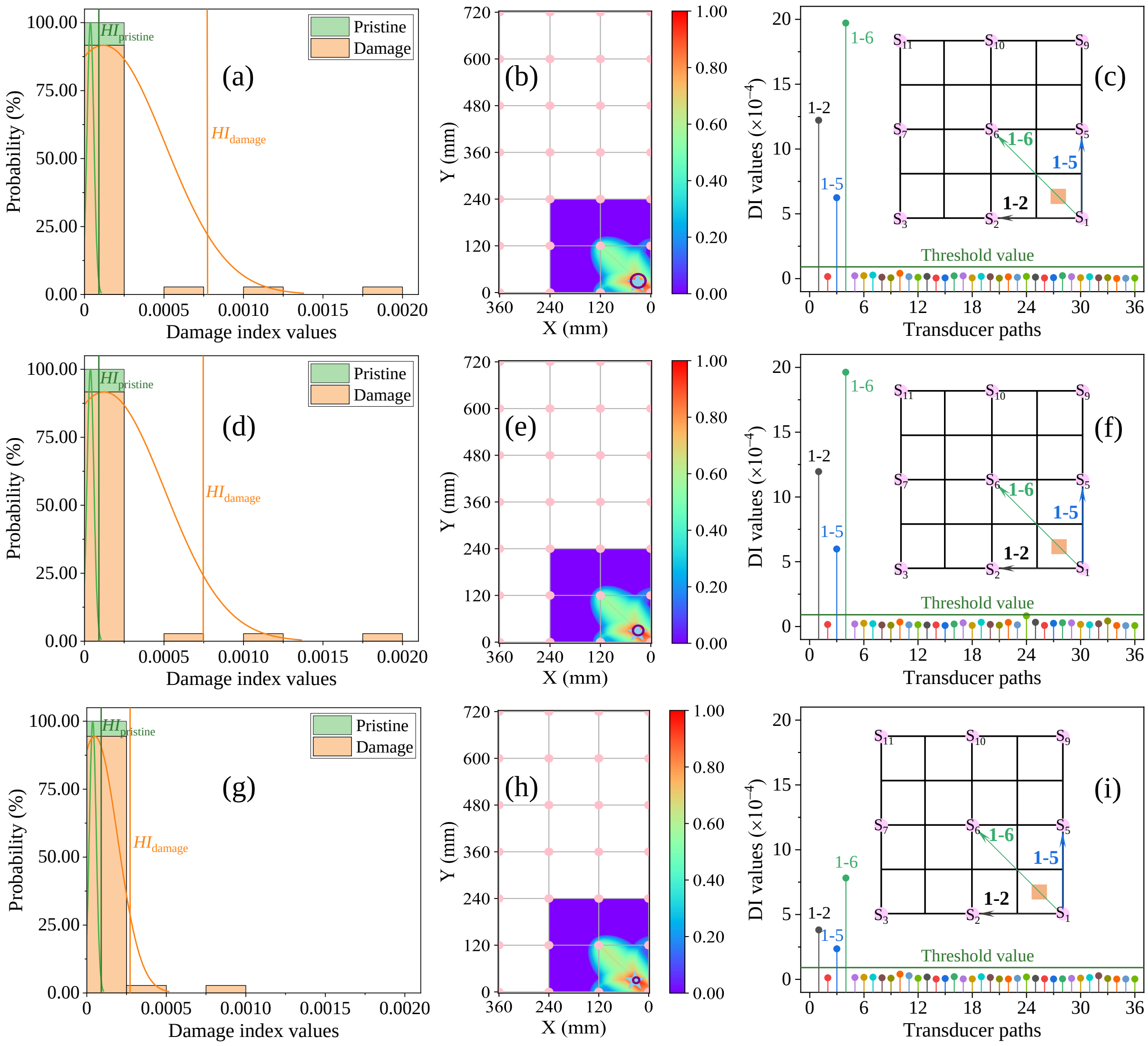

Fig. 10. **Damage detection and localization results for a single damage under the damage diameters of 35 mm, 25 mm, and 15 mm, respectively (from top to bottom)**: (a, d, g) DI distributions; (b, e, h) Damage imaging maps; and (c, f, i) Histograms of DI ignoring the inverse and indirect paths.

Furthermore, the performance of the proposed method for damage detection and localization in different regions was further investigated. Fig. 11(a, d, g) presents the distribution of DI values in the region where the damage is located for different regions. It can be seen that for the single damage case, the distribution of DI under different regions produces a similar trend, but with slightly different $HI_{damage}$. Fig. 11(b, e, h) demonstrates the damage imaging results of the proposed method in different regions, and it is clear that the proposed method can effectively localize the damage. Fig. 11(c, f, i) shows the distribution of DI values in the damage region, and it can be observed that the sensing paths with larger DI values are all located in the vicinity of the damage, and all of them are accurately identified by the proposed model. The above results show that the proposed method for baseline-free damage detection and localization has good generalization and scalability.

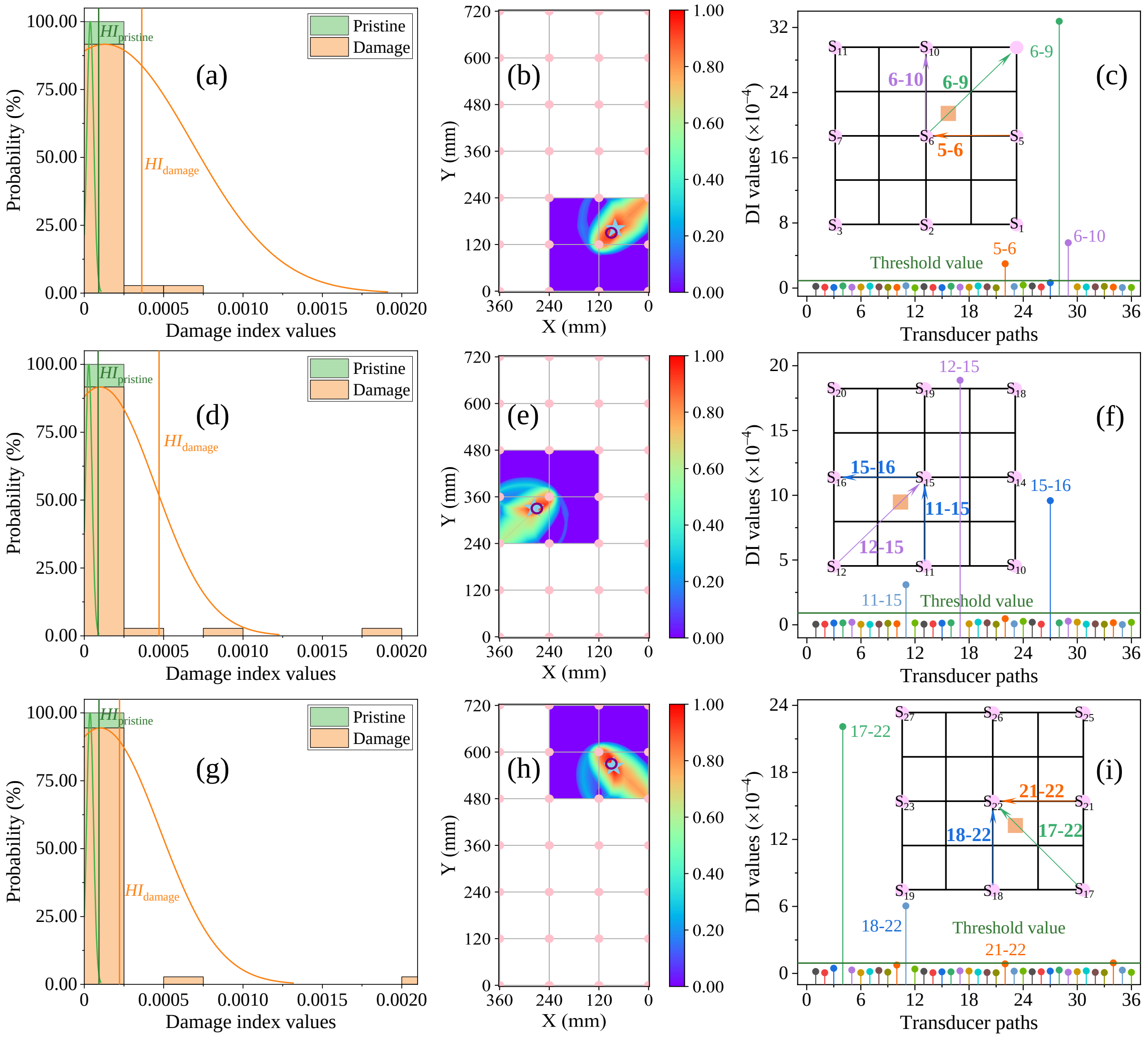

Fig. 11. **Damage detection and localization results for a single damage under different target regions (from top to bottom)**: (a, d, g) DI distributions; (b, e, h) Damage imaging maps; and (c, f, i) Histograms of DI ignoring the inverse and indirect paths.

### 3.1.3. The results of virtual baseline signal reconstruction

The reconstruction of the virtual baseline signals is realized with the KAE according to Section 2.1. Due to space limitations, Fig. 12 shows an example of reconstruction of the signals of some of the sensing paths where the damage is located at (30 mm, 30 mm) (consistent with the damage location of Fig. 9). According to Fig. 9, sensing paths A1-S2, A1-S5, and A1-S6 are identified as damaged paths, and the remaining paths are considered to be undamaged paths. Fig. 12(a-d) display reconstruction results for the damaged paths, showing significant deviations from the baseline signals, including reduced amplitude and phase delays, highlighting an inability to accurately replicate the baseline. In contrast, Fig. 12(e, f) present results for undamaged paths, where reconstructed signals closely match the experimental baseline, confirming their undamaged state and aligning with Fig. 9. These findings demonstrate the strong capability and exceptional performance of KAE in reconstructing virtual baseline signals for SHM.

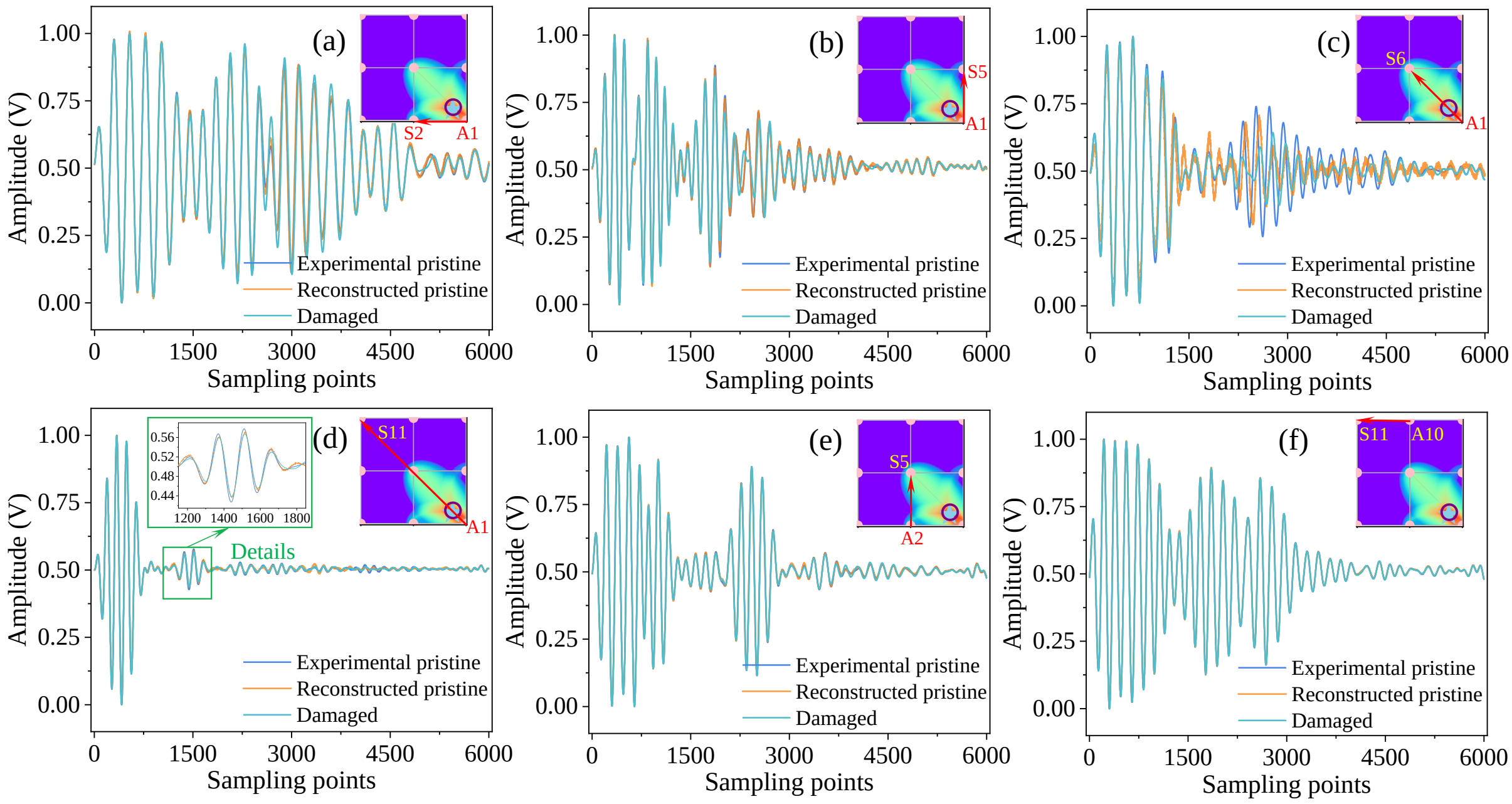

Fig. 12. **Reconstruction results of the signals from different sensing paths when the damage is located at (30, 30)**: (a) A1-S2; (b) A1-S5; (c) A1-S6; (d) A1-S11; (e) A2-S5; (f) A10-S11. (Details of the sensing path and damage are shown in the upper right corner of each subfigure).

### 3.1.4. Multiple damage detection and localization

Fig. 13 illustrates the damage detection and localization results for two damage scenarios using the proposed baseline-free algorithm. Fig. 13(a)-(f) show the progression of $HI_{\text{damage}}$ values in various target regions. $HI_{\text{damage}}$ values in damaged regions are significantly higher than $HI_{\text{pristine}}$ (Fig. 13(b, c)), while intact regions maintain $HI_{\text{damage}}$ values below $HI_{\text{pristine}}$. This clear distinction demonstrates the method's high sensitivity and reliability in detecting and differentiating damaged regions. Fig. 13(g) shows the performance of the proposed model in multi-damage localization, and the results show that the proposed method can effectively localize the location of the damage. Fig. 13(h) shows the distribution of DI values for all sensing paths throughout the monitored area. It can be observed that the method successfully captures the extreme DI values presented by the sensing paths close to the damage and that $HI_{\text{damage}}$ is always greater than $HI_{\text{pristine}}$ in the region where the damage occurs and vice versa.

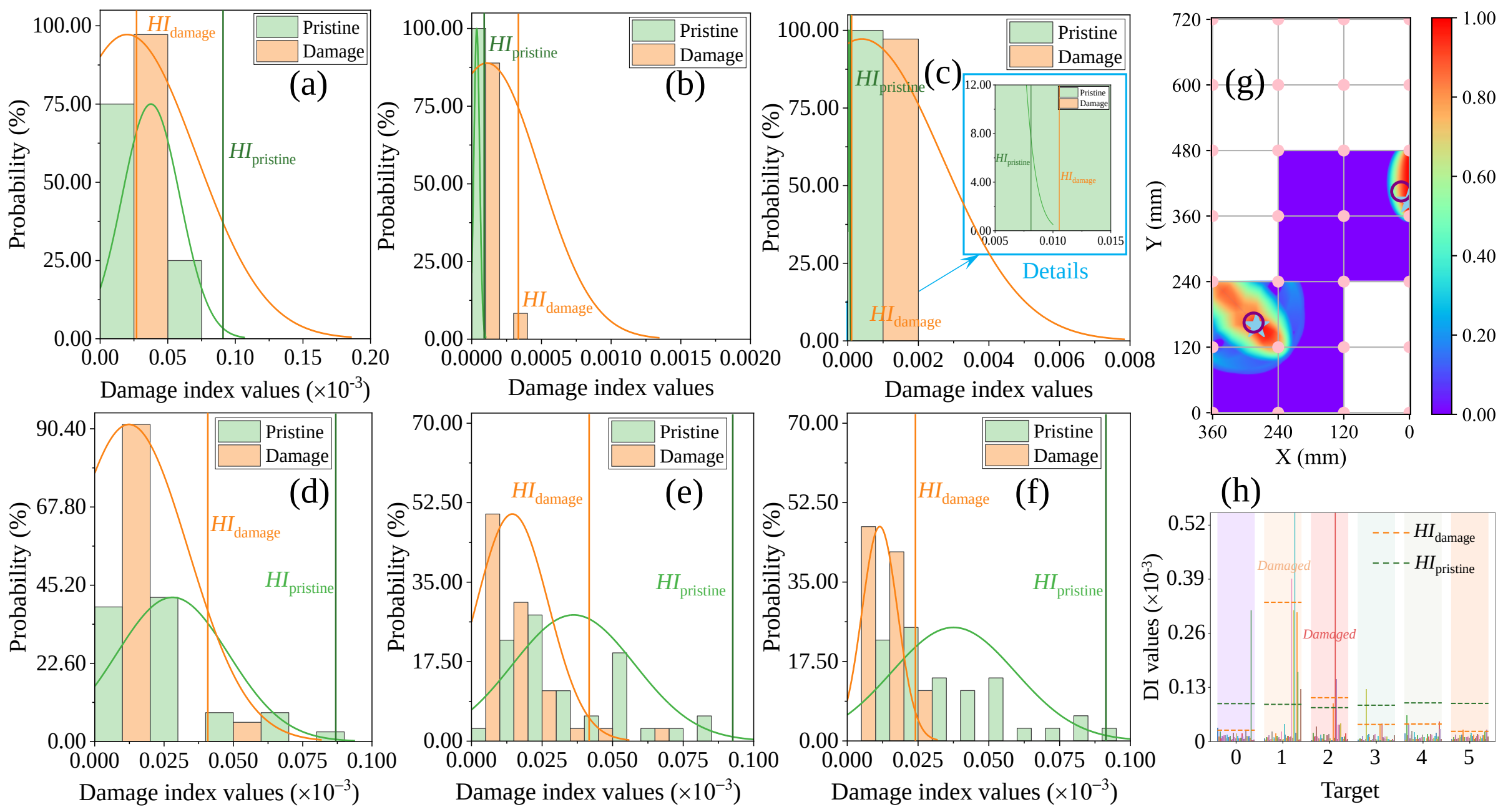

Fig. 13. **(a)-(f) DI distribution in different target regions**: (a) target 0, (b) target 1, (c) target 2, (d) target 3, (e) target 4, (f) target 5; (g) Damage imaging map; (h) DI histograms with reciprocal paths and indirect paths ignored.

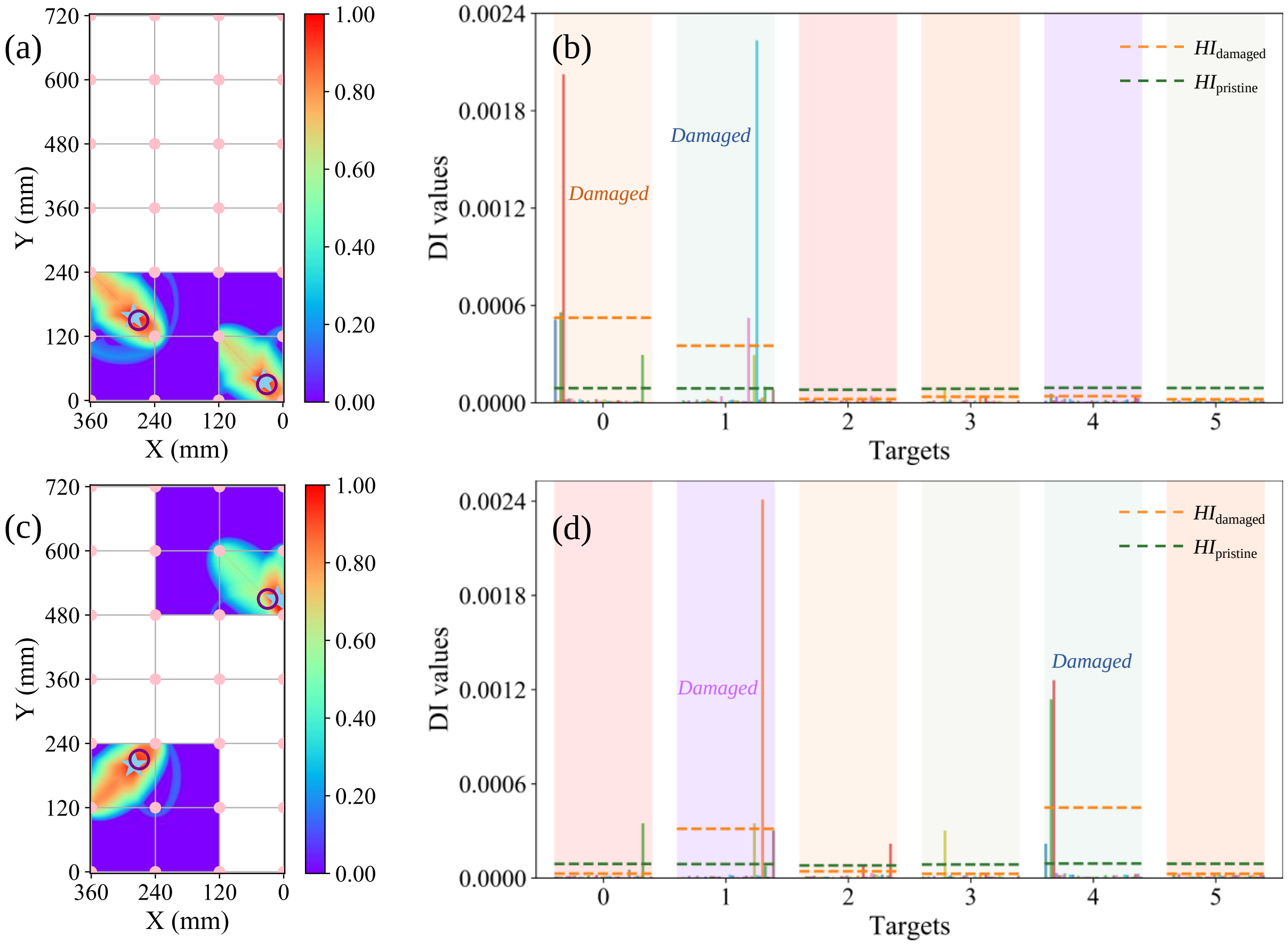

Fig. 14. **Damage detection and localization results for two damages**: (a, c) Damage imaging maps; (b, d) DI histograms and detection threshold maps.

Fig. 14 illustrates the damage detection and localization performance in scenarios involving two simultaneous damages. The method successfully identifies and accurately localizes both damages. In Fig. 14(b, d), the DI values and corresponding $HI_{\text{damage}}$ and $HI_{\text{pristine}}$ for different target regions are presented. The results indicate that the $HI_{\text{damage}}$ in the damaged regions is consistently larger than the $HI_{\text{pristine}}$, highlighting a significant contrast between damaged and normal regions. This validates the method's effectiveness in dual-damage detection, as it not only identifies the damaged regions but also accurately localizes the damages within those regions.

#### 3.1.5. Discussion of damage localization with reduced number of sensing paths

Currently, the performance of the proposed baseline-free damage localization method for damage detection and localization has been evaluated using all available sensing paths (number 36). However, the previous experiments revealed that only a few sensing paths exhibit a large DI value (See Fig. 9-Fig. 11). Therefore, a novel self-selecting path algorithm is proposed in this section. Specifically, only the paths with large DIs among the sensing paths are used for damage imaging. Table 4 demonstrates the proposed baseline-free damage localization method for different numbers of sensing paths. It can be seen that even with a limited number of sensing paths, the proposed baseline-free damage localization method exhibits excellent damage localization performance.

Table 4. **Performance comparison with different number of sensing paths**

| Path number | RMSE (mm) | MAPE | MRE | Time (s) |
|---|---|---|---|---|
| 3 | 14.34 | 6.23% | 5.61% | 50.13 |
| 9 | 13.97 | 5.93% | 5.50% | 57.93 |
| 15 | 13.90 | 5.91% | 5.45% | 69.13 |
| 21 | 13.97 | 5.93% | 5.50% | 80.61 |
| 27 | 13.90 | 5.91% | 5.45% | 90.23 |
| 33 | 14.79 | 7.29% | 5.96% | 102.35 |

#### 3.1.6. Comparison with other state-of-the-art methods

To further validate the superiority of the proposed method, comparisons were made with the state-of-the-art probabilistic imaging algorithms. The comparison techniques are CAE-RAPID [22], modified RAPID [30], and RAPDI [34], among which CAE-RAPID is an excellent unsupervised learning-based baseline-free damage localization algorithm. For the proposed methods, comparisons were made using GW data from the same damage location, respectively. The structure and parameters of the compared methods were set to similar values as those reported in their original studies and optimized to give the best damage localization performance. The localization results of the above four methods under randomly selected damages are shown in Table 5. It can be clearly seen that neither the traditional RAPID nor CAE-RAPID can locate all the damages with high accuracy, while the newly developed KAE-MRAPID can obtain accurate localization results, the lowest values for RMSE (14.20 mm) and MRE (5.51%), respectively. Furthermore, by employing self-selecting paths, the method significantly reduces the time required for damage detection and localization, with the entire process completed in 66.19 seconds. These findings underscore the method's high computational efficiency while maintaining competitive localization accuracy, making it an effective and practical solution for damage detection and localization tasks.

The key mysteries that make KAE-MRAPID superior to other methods are twofold: First, the proposed KAE autoencoder can extract more robust DI values. Second, the MRAPID algorithm can obtain more accurate weight values for damage distribution.

Table 5. Comparison of localization accuracy between different damage imaging algorithms

| No. | Damage center | RAPID | Modified RAPID | CAE-RAPID | Proposed method |
|---|---|---|---|---|---|
| 1 | (90, 150) | (120, 120) | (78.8, 161.8) | (80.6, 143.0) | (80.8, 161.0) |
| 2 | (270, 330) | (240, 360) | (279.8, 323.4) | (259.0, 346.4) | (276.8, 331.2) |
| 3 | (90, 570) | (120, 600) | (100, 590.0) | (112.8, 546.2) | (84.4, 563.6) |
| 4 | (30, 30) | (0, 0) | (38.2, 38.6) | (36.2, 36.2) | (33, 30.4) |
| 5 | (210, 150) | (240, 120) | (239.8, 138.4) | (241.0, 119.80) | (240.8, 141.8) |
| 6 | (150, 30) | (120, 0) | (161.6, 40.2) | (156.6, 36.0) | (157, 32.6) |
| 7 | (210, 90) | (240, 120) | (208.0, 83.0) | (190.6, 120.0) | (209.6, 91.8) |
| 8 | (90, 90) | (120, 120) | (88.0, 81.4) | (106.4, 101.0) | (103.8, 93.6) |
| RMSE | / | 42.43 mm | 17.46 mm | 27.87 mm | 14.2 mm |
| MRE | / | 29.70% | 9.57% | 12.03% | 5.51% |
| Time | / | 60.59 s | 109.22 s | 53.72 s | 66.19 s |

### 3.1.7. Verifying the superiority of the proposed KAE

To validate the superiority of the proposed KAE model over convolutional autoencoder in baseline-free damage localization, two models are considered as follows: (1) KAE-RAPID: Replacing the convolutional autoencoder in CAE-RAPID [22] with KAE while keeping all other parameters unchanged. (2) CAE-MRAPID: Replacing the KAE in the proposed KAE-MRAPID with a convolutional autoencoder while maintaining identical parameters.

The performance of these two models was tested using randomly selected damage locations from Table 5. Table 6 presents the comparative localization performance results for the different models. It can be observed that replacing the convolutional autoencoder in CAE-RAPID with KAE significantly reduces the RMSE, MAPE, and MRE of the model, indicating improved localization performance. However, replacing KAE in KAE-MRAPID with a convolutional autoencoder increases the RMSE from 14.2 mm to 25.43 mm, MAPE from 5.96% to 13.71%, and MRE from 5.51% to 11.92%. This suggests a substantial degradation in localization performance. The results demonstrate that compared to convolutional autoencoder, KAE can better reconstruct virtual baseline signals and fully extract damage information from each sensing path. It provides more robust damage indexes for the damage imaging method, enabling more accurate damage localization.

Table 6. Damage localization performance with different localization methods

| Method | RMSE (mm) | MAPE | MRE |
|---|---|---|---|
| CAE-RAPID | 27.87 | 11.99% | 12.03% |
| KAE-RAPID | 26.41 | 11.26% | 11.33% |
| KAE-MRAPID | 14.2 | 5.96% | 5.51% |
| CAE-MRAPID | 25.43 | 13.71% | 11.92% |

### 3.2. Case 2: Application to the circular through-hole composites

To further validate the effectiveness and robustness of the proposed baseline-free damage detection and localization technique, actual damage assessment (i.e., circular through-hole) was carried out on a composite panel with dimensions of 450 mm × 450 mm × 3 mm, as shown in Fig. 15(a). Experiments were conducted at different locations (① (127,185), (② (147, 245), ③ (320, 245), and ④ (340, 185)) for four damage scenarios. The diameter of all four through-hole damages was 12 mm, and these holes were drilled with a drill. Similar to Section 3.1, 16 PZT-5A were used as actuators or sensors and cured on composite panels using an AB-type epoxy adhesive. The PZT have a diameter of 8 mm and a thickness of 0.48 mm. A total of 78 sensing paths were constructed in the whole piezoelectric sensing network, and some typical sensing paths are shown in Fig. 15(b).

During the experiment, a five-cycle Hanning window signal with a centre frequency of 80 kHz was excited and received at a sampling rate of 12 MHz. The peak-to-peak voltage and gain of the output signal were set to 60 V and 20 dB, respectively. The response of the associated GW signal was recorded by the ultrasound detector mentioned in Section 3.1. The GW signals of the baseline and damage signals (before and after the introduction of the hole damage propagation) were captured in the experiments. More details about the setup and technical content of the experiment can be found in [35].

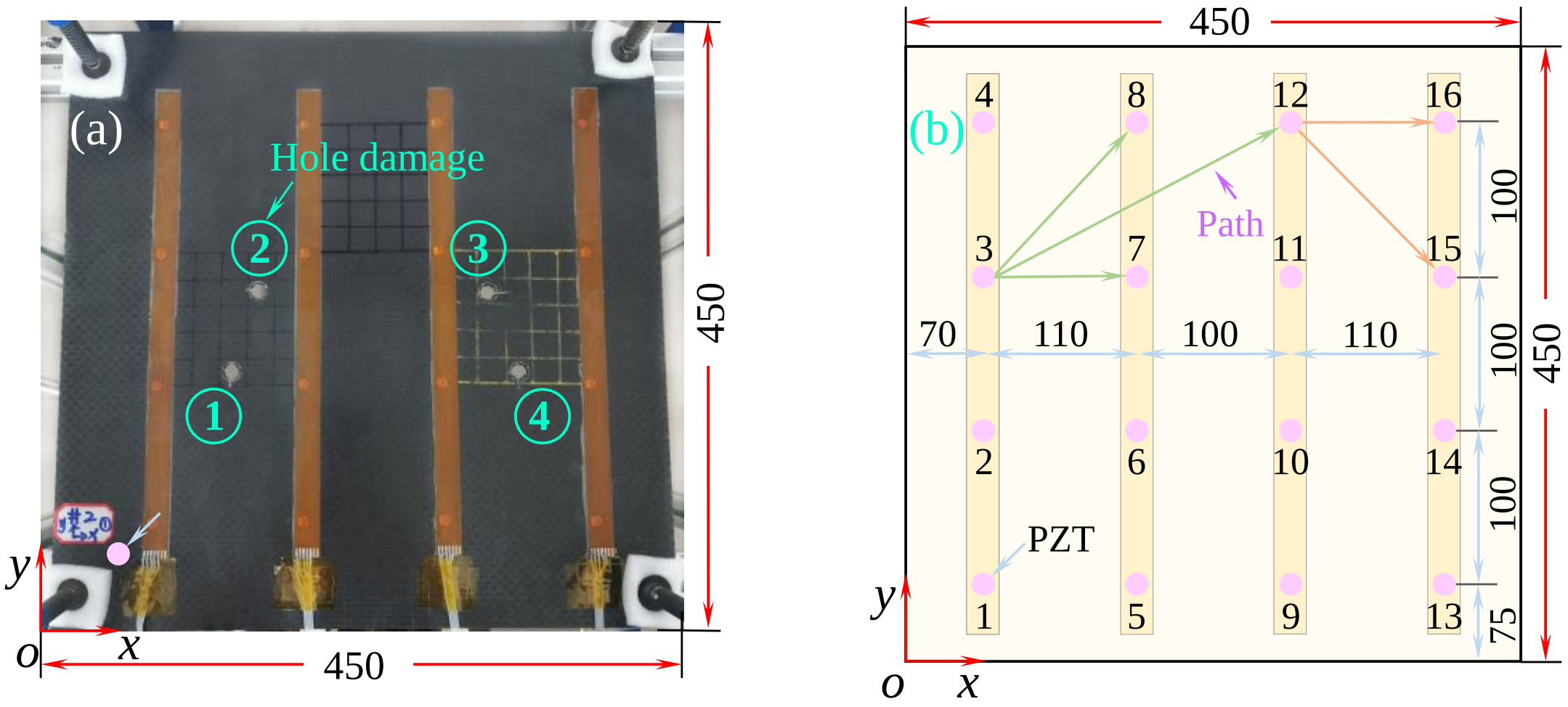

Fig. 15. **PZT sensors layout**: (a) composite flat structures with four through-hole damages; (b) Schematic diagram of the sensor layout and sensing paths.

### 3.2.1. Parameter setting and training process

To further validate the generality of the proposed baseline-free damage detection and localization method, the internal parameters of the KAE used in the flat plate structure are identical to those set in the wind turbine blade model, except for the batch size and the epochs. In the flat plate structure, the batch size of the KAE is 4 and the number of epochs is 200. This is mainly due to the fact that the baseline signal acquired in the flat structure is much less compared to the baseline data acquired on the wind turbine blade. The learning curves of the proposed KAE model on raw GW signals from composite flat structures is shown in Fig. 16. The performance of KAE is similar to that shown in Fig. 6, and the description is not repeated here due to page limitations.

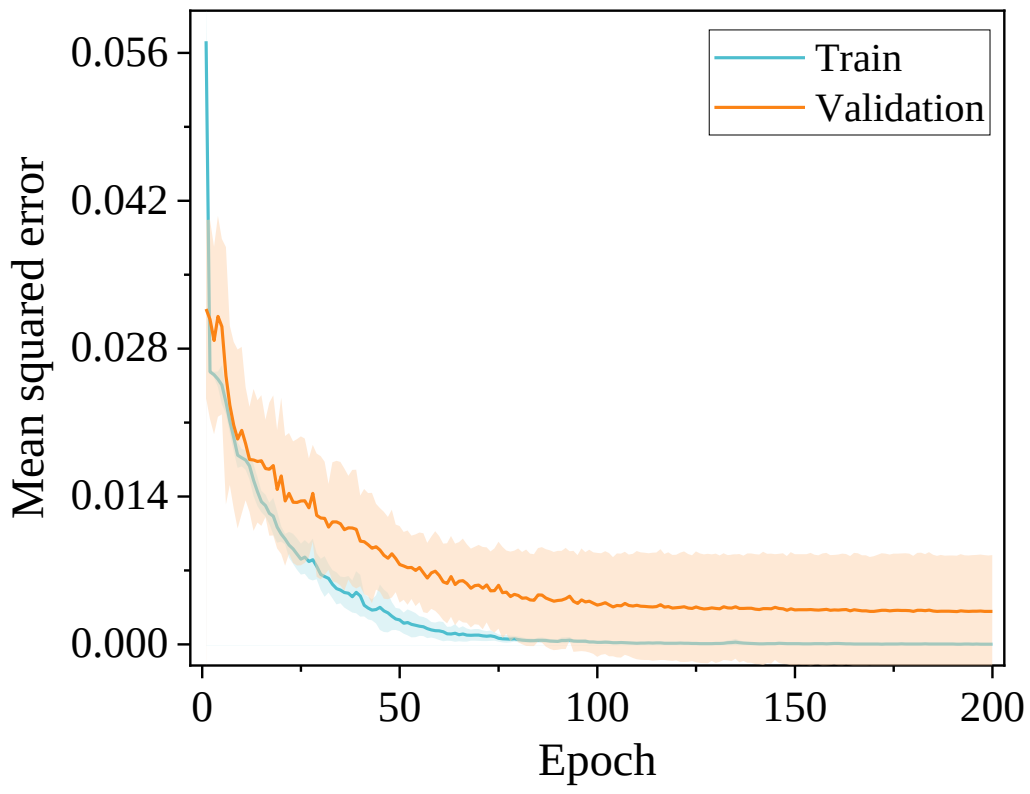

Fig. 16. **Learning curves of the proposed KAE on raw GW signals from composite flat structures.**

### 3.2.2. Damage detection and localization

Fig. 17(a, d, g, j) presents the evolution of the *HI* values following damage initiation in the monitored area. Notably, the $HI_{\text{damage}}$ values consistently exceed the $HI_{\text{pristine}}$ values by a significant margin. Fig. 17(b, e, h, k) demonstrates the damage detection and localization results obtained using the proposed baseline-free detection algorithm. The results indicate that the proposed method achieves accurate damage localization. Fig. 17(c, f, i, l) displays the DI values across all sensing paths, revealing that the proposed method effectively identifies the extreme DI values adjacent to the damage paths and successfully utilizes these paths for precise damage localization.

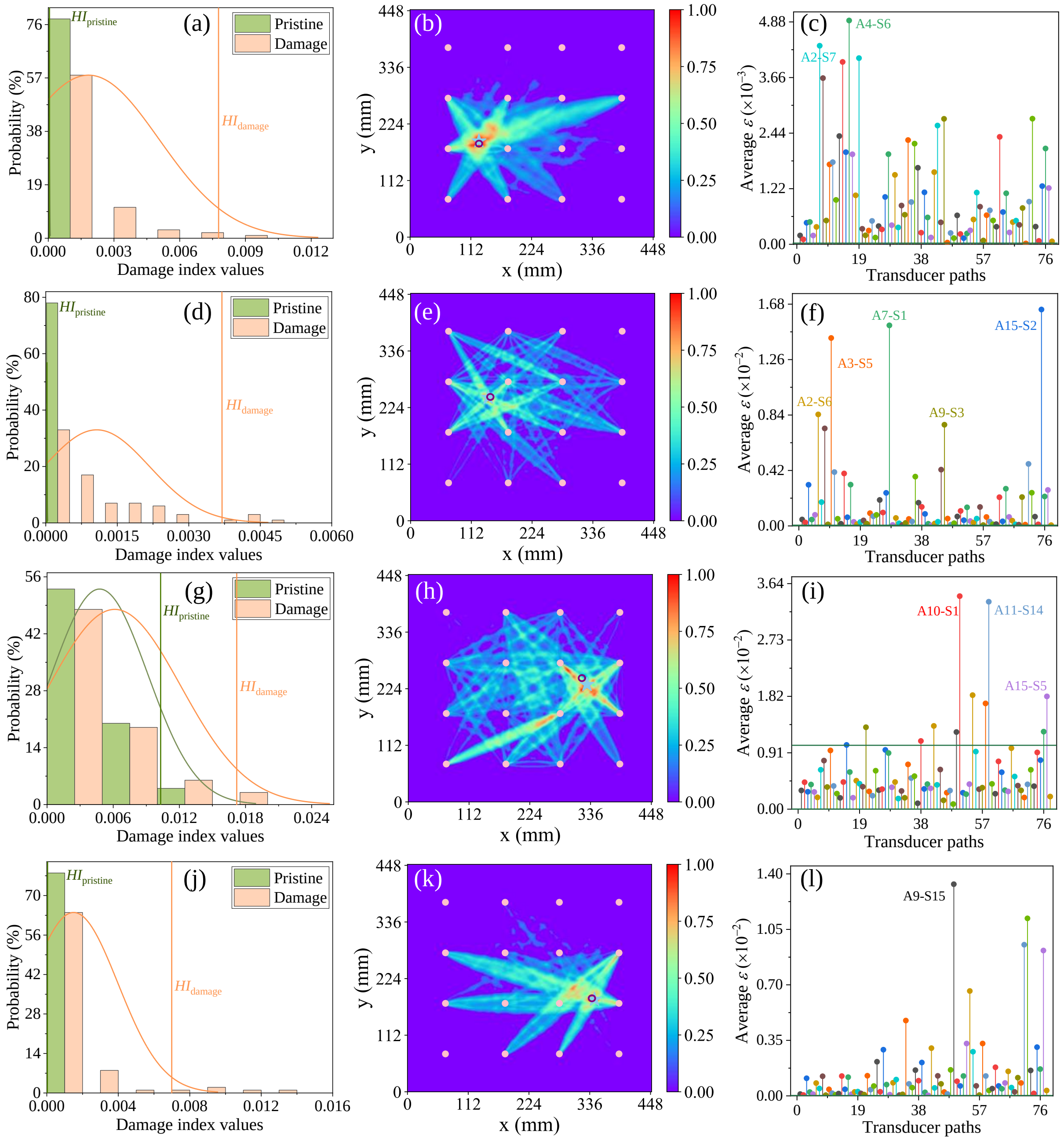

Fig. 17. **Damage detection and localization results**: (a, d, g, j) DI distributions; (b, e, h, k) Damage imaging map for four through-hole damages; (c, f, i, l) Histograms of DI ignoring the inverse and indirect paths. The actual damage is marked with a purple circle, and the predicted damage is represented with a sky blue star.

## 4. Conclusions

In this paper, a hybrid framework combining the unsupervised KAE and the MRAPID algorithm is proposed for baseline-free damage detection and localization within monitored structures. The proposed KAE-MRAPID model achieves damage detection and localization in a fully unsupervised and end-to-end manner. It eliminates the need for prior knowledge of material properties (such as wave speed, or the effects of damage on amplitude and wave mode phase) for preprocessing GW signals and manually extracting features. Therefore, the proposed method can be effectively extended to GW-based SHM of any structure that supports GW propagation.

The performance of the method was validated using simulated damage data obtained on wind turbine blades and actual damage data obtained on composite panels. Experimental results demonstrate that the proposed fully unsupervised learning-based baseline-free damage detection and localization model outperform the traditional RAPID algorithms and the currently reported unsupervised learning-based baseline-free damage detection and localization methods. Specifically, under the same data conditions, the MRE of traditional RAPID, modified RAPID, and CAE-RAPID were 29.70%, 9.57%, and 12.03%, respectively. In contrast, the proposed KAE-MRAPID achieved an MRE of only 5.51%, demonstrating the best localization performance. Furthermore, the proposed KAE-MRAPID supports the detection and localization for multiple damages, which is not currently supported by other methods, representing another advantage of our algorithm.

It should be noted that although the proposed hybrid KAE and MRAPID algorithm-based baseline-free damage detection and localization method provided more accurate and conservative predictions and can detect and locate damage in a baseline-free manner, it is not suitable for damage quantification. This imperfection highlights a key area for future research and resolution. Moreover, considering the effect of temperature on the GW signals is also the focus of our next research.

## Declaration of competing interest

The authors declare that they have no known competing financial interests or personal relationships that could have appeared to influence the work reported in this paper.

## Data availability

The authors do not have permission to share data.

## Acknowledgments

This work was supported by National Natural Science Foundation of China (under Grant

No. U2141245, 52405148). The authors thank the anonymous reviewers for their valuable suggestions to improve this paper. Yunlai Liao acknowledges the support by China Scholarship Council.